\documentclass[10pt,letterpaper,twocolumn]{article} 

\usepackage{bm,braket}
\usepackage{float}
\usepackage{graphicx}
\usepackage{amsfonts}
\usepackage{amsmath}
\usepackage{amssymb}
\usepackage{color}
\usepackage{ulem}
\usepackage{overpic}
\usepackage{ol2}
\newcommand {\be}{\begin{equation}}
\newcommand {\ee}{\end{equation}}
\newcommand {\bea}{\begin{eqnarray}}
\newcommand {\eea}{\end{eqnarray}}

\renewcommand{\v}[1]{\ensuremath{\mathbf{#1}}} 

\def\r1{\textbf{r}}

\begin{document}

\twocolumn[
\abovedisplayskip=6pt
\abovedisplayshortskip=6pt
\belowdisplayskip=6pt
\belowdisplayshortskip=6pt

\title{Quasi-normal mode approach to the local-field
problem in quantum optics}

\author{Rong-Chun Ge$^{1,*}$, Jeff F. Young$^2$,  and S. Hughes$^1$}
\address{
$^1$Department of Physics, Engineering Physics and Astronomy,
Queen's University, Kingston, ON, Canada K7L 3N6\\
$^2$ Department of Physics and Astronomy, University of British Columbia, 6224 Agricultural
Road, Vancouver, BC,  Canada V6T 1Z1\\
$^*$Corresponding author: rchge@physics.queensu.ca
}





\begin{abstract}
The local-field (LF) problem of a finite-size dipole emit-
ter radiating inside a lossy inhomogeneous structure
is a long-standing and challenging quantum optical
problem, and it now is becoming more important due
to rapid advances in solid-state fabrication technologies.
Here we introduce a simple and accurate quasi-normal
mode (QNM) technique to solve this problem analyti-
cally by separating the scattering problem into contribu-
tions from the QNM and an image dipole. Using a real-
cavity model to describe an artificial atom inside a lossy
and dispersive gold nanorod, we show when the contri-
bution of the QNM to LFs will dominate over the homo-
geneous contribution. We also show how to accurately
describe surface scattering for real cavities that are close
to the metal interface and explore regimes when the
surface scattering dominates. Our results offer an intui-
tive picture of the underlying physics for the LF problem
and facilitate the understanding of novel photon sources
within lossy structures.
\end{abstract}

]

The local-field (LF) problem in quantum optics has a long,
rich history and is now becoming more important due to
the development of advanced nanofabrication technologies
for optical structures and nanophotonic devices. As is well
known, the local photon density of states (LDOS) largely controls light-matter interactions, including the spontaneous
emission (SE) of a dipole emitter~\cite{qdinphotonicrystal} located in an arbitrary
scattering environment. The LDOS is related to the photon
Green function (GF) of the medium, through ${\rm LDOS} \propto 
{\rm
Im}[{\bf G}({\bf r},{\bf
r}'={\bf r};\omega)]$~\cite{nanoptics},
where ${\bf G}({\bf r},{\bf
r}';\omega)$ is the electric-field response
at ${\bf r}$ from a dipole emitter at ${\bf r}'$.
For example, the Im(${\bf G}$) inside a homogeneous lossless dielectric is simply the
vacuum LDOS scaled by the refractive index of the dielectric.
A plethora of novel linear and nonlinear optical effects have
been predicted and demonstrated in nanophotonic structures
engineered to have an LDOS that differs substantially from
that in uniform lossless media~\cite{plasmonBook}.  The vast majority of the
experimentally verified LDOS effects have been obtained
when dipole emitters are located in or in proximity to lossless
dielectric structures, or in proximity to lossy (metallic) struc-
tures. In all of these cases, the Im(${\bf G}$) is well-behaved and there is typically good agreement between experimental results of the modified LDOS and model calculations using Im(${\bf G}$).
  Rapid advances in experimental nanoplasmonics mean that it is now possible to embed  dipole emitters {\it within} inhomogeneous 
metallic 
microcavities~\cite{Suemene11,Babinec14}.
This raises the challenging theoretical problem of how to deal with the fact that the
LDOS is infinity inside lossy, dispersive structures for which the imaginary part of the permittivity is
non-zero,i.e., ${\rm Im}[\varepsilon(\omega)]\neq0$.

Related theoretical work was motivated by dispersive lossy materials in which dipole-emitting atoms or ions might be interstitially or substitutionally located.
 Two distinct models were developed, both of which isolated the dipole ``defect'' within a cavity that occupied some volume in the vacuum between host atoms.  The virtual cavity (VC) and real cavity (RC) models differed in how the LF is treated~\cite{local1}: in the VC model, the LF is obtained by averaging the macroscopic field inside the cavity, whereas in the RC model, the LF is obtained by self-consistently including the scattering from the cavity boundary.
Relevant experiments, though limited, suggest that for ion-doped lossy materials, the RC model works well~\cite{local3exp}, however comparison between experiment and model relies on fitting the effective "cavity radius" of the ion, a rather ill-defined physical quantity.
A significant practical and conceptual advantage of nanofabricated structures that surround finite-size nano-dipoles~\cite{Suemene11,Babinec14,Reuven15} is that there is typically no ambiguity associated with the size and shape of the RC in which the dipole is located, and it is manifestly clear that the RC, versus the VC model should apply when dealing with the all-important LF problem.

Most works on the RC model to date deal with the LF inside a lossy {\it homogeneous} material.
It has been shown the Born approximation
could be used to address the arguably more important LF problem in an {\it inhomogeneous} lossy
structure~\cite{finitelocal}, though it has been confined to spherical
cavity structures.
Recently, the LF effect in the strong coupling regime for
an atom in a spherical structure has also been studied~\cite{stronglocal}.
For non-spherical geometries, the finite-difference time-domain (FDTD)~\cite{finitefdtd}
algorithm is effective for computing the LF; however, the simulations are very computational-time expensive and it is difficult to understand the underlying physics of the scattering problem. Thus one would like a more efficient theoretical approach to describing the SE rate from emitters located inside arbitrarily shaped, lossy material structures.  It would be especially useful to have a formalism that facilitates quick exploration of parameter space, e.g.,  to explore  the role of RC size and position, on the overall device performance.  Such models are also relevant to the study of
 spasers~\cite{spasertheory,spaserexperiment}
 and quantum plasmonics~\cite{quantumplasmonics}.

In this Letter, we introduce a QNM technique
developed for metallic nanoresonator (MNRs)~\cite{modesandvolumes,quasius1,quasius2,quasi3}
to address the LF problem with a RC model~wherein the radius of the RC is directly associated with the volume of the emitter,  e,g., an artificial atom or quantum dot (QD).
 The MNR problem is particularly challenging from a numerical perspective because of the large complex permittivity, requiring very small (sub-nm) computational gridding and tediously long simulation times. However, we show that
the LF can be solved semi-analytically for lossy cavity structures,
even for complex-shaped MNRs, if one adopts a QNM approach.
The QNMs are the discrete modes of an open system,
and localized surface plasmons (LSPs) may be directly understood as QNMs of the MNRs~\cite{modesandvolumes,quasi3},
defined as the frequency domain solutions to the wave equation with open boundary conditions~\cite{quasi1,quasi2}.
The QNM has been  widely exploited in various
field of physics~\cite{quasi1,quasi2}, and a QNM approach is particularly useful
when the response of the system is dominated by one or several
discrete resonances. In this work, we study the example of a RC inside  a gold nanorod.
To check the accuracy of our theory, we first specify a RC embedded at the center of the nanorod,
and compute the SE enhancement factor using
an established FDTD technique~\cite{finitefdtd} and  our semi-analytical method. We find an excellent agreement between both methods for the LF computation.  Our approach not only offers orders-of-magnitude improvement in the
required computational time, but  helps to identify the
underlying physics in an intuitive and easy-to-understand way.
Using this semi-analytic technique, we then investigate the role of emitter size, and find that, for large RC emitters
such as a QDs, the contribution of the QNM can
dominate over the homogeneous contribution of the gold. In addition,
we also explore the SE rate for RCs placed near
the MNR surface (inside the lossy structure), and show that, ignoring quantum tunnelling effects, the physics can be accurately captured using an image dipole technique. We again explore regimes when the surface scattering
can be the dominant mechanism for SE enhancement.

\begin{figure}[t]
  \centering
   \begin{overpic}[width=0.45\textwidth]{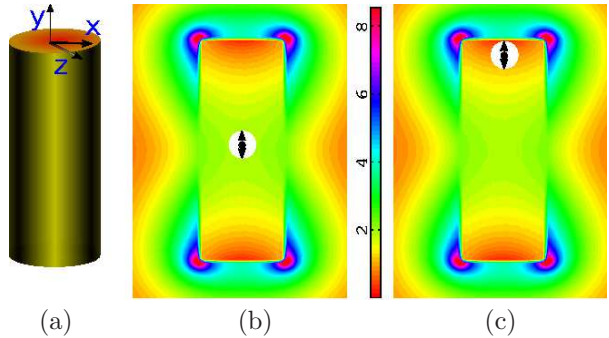}
   \put(5,-5){(a)}
\put(38,-5){(b)}
\put(79,-5){(c)}
   \end{overpic}
   \vspace{0.2cm}
     \caption{\footnotesize{(a) Schematic  of a 3D gold nanorod in a homogeneous background ($n_b=1.5$). (b-c)  Spatial profile of the QNM, $\tilde {\bf f}_c$  with $\omega_c/2\pi = 324.981~$THz (1.344 eV); (b) indicates  the LF problem with a RC near the center of the nanorod along the $y$-axis, and (c)  shows the same RC but $1.5~$nm away from the top surface. The radius and permittivity of the cavity are $r_c$ and $\varepsilon_c$, respectively (see text).}}
   \vspace{-0.1cm}
\label{f1}
\end{figure}

In the following we consider a 3D gold nanorod as our
stereotypical
 arbitrarily-shaped lossy material. Gold MNRs
can achieve LSP responses in the visible. Other geometries including
patch antennas have experimentally been probed to achieve enhanced SE of QDs~\cite{opticalantennas}. 
Figure~\ref{f1}(a) shows a schematic  of the gold rod with a length  100~nm, radius $r_a 
=15~$nm, lying along $y$-axis with a background index $n_b = 1.5$
($\varepsilon_b = n_b^2$). For the gold MNR,  we use the Drude model,
$\varepsilon(\omega) = 1-\frac{\omega_p^2}{\omega^2+i\omega\gamma}$
with parameters: $\omega_p = 1.26 \times 10^{16}~$rad/s, $\gamma = 1.41\times
10^{14}~$rad/s.

In Ref.~\cite{local2}, Toma\v{s} showed that the LF could be
described by a combination of both the volume averaged homogeneous contribution (using the homogeneous GF for gold)  ${\bf G}_{\rm loc}^{\rm h}({\bf r}_d, {\bf r}_d;\omega)$ (${\bf r}_d$ is the position of an electric dipole at the center of the RC) and  the scattering contribution, ${\bf G}_{\rm loc}^{\rm sc}({\bf r}_d, {\bf r}_d;\omega)$, due to the inhomogeneity of the lossy structure; so the LF GF is given by~\cite{local2}
\begin{align}
{\bf G}_{\rm loc}({\bf r}_d, {\bf r}_d;\omega) = {\bf G}_{\rm loc}^{\rm h}({\bf r}_d, {\bf r}_d;\omega) + {\bf G}_{\rm loc}^{\rm sc}({\bf r}_d, {\bf r}_d;\omega),
\label{local0}
\end{align}
where ${\bf G}_{\rm loc}^{\rm sc}({\bf r}_d, {\bf r}_d;\omega)=\left(\frac{3\varepsilon}{2\varepsilon+\varepsilon_c}\right)^2{\bf G}^{\rm sc}({\bf r}_d, {\bf r}_d;\omega)$ and ${\bf G}^{\rm sc}({\bf r}_d, {\bf r}_d;\omega)$ is the scattered GF without the RC.
Using a Born-expansion method, Dung {\it et al}.,~\cite{finitelocal}
showed direct agreement with the Toma\v{s} LF formula.
The homogeneous contribution is obtained analytically with the two-layer spherical model,
yielding a $1/r_c^3$ dependence at a small cavity radius: ${\rm Im}[\hat{n}\cdot{\bf G}_{\rm loc}^{h}\cdot\hat{n}]\approx \frac{9{\rm Im}[\varepsilon]}{|2\varepsilon+\varepsilon_c|^2}\frac{1}{6\pi r_c^3}$.
Thus one recognizes that the LF problem is essentially solved if one can obtain ${\bf G}^{\rm sc}_{\rm loc}$, though this is far from trivial, except for very simple
shapes.

With regards to MNRs,
it has been shown the scattering behaviour can be accurately
 described in terms of the QNMs~\cite{quasius1,quasius2,quasi3}, and a mode
expansion technique for the GF has been developed and
confirmed to be very accurate~\cite{quasius1,quasius2}. The
QNMs of the  system, $\tilde{\bf f}_{\mu}$, have
complex eigenfrequencies $\tilde{\omega}_\mu = \omega_\mu - {\rm
i}\gamma_\mu$, and are normalized as follows:~\cite{quasi1,Lee99}
\begin{align}
\langle\langle \tilde{\bf f}_{\mu}|\tilde{\bf f}_{\nu}\rangle\rangle\!&=\!\lim_{V\rightarrow\infty}\int_V\left(\frac{1}{2\omega}\frac{\partial (\varepsilon({\bf r},\omega)\omega^2)}{\partial \omega}\right)_{\omega=\tilde{\omega}_{\mu}}\!\!\!\!\tilde{\bf f}_{\mu}({\bf r})\cdot\tilde{\bf f}_{\nu}({\bf r})d{\bf r} \nonumber\\
&+ \frac{ic}{2\tilde{\omega}_{\mu}}\int_{\partial V}\sqrt{\varepsilon({\bf r})}\tilde{\bf f}_{\mu}({\bf r})\cdot\tilde{\bf f}_{\nu}({\bf r})d{\bf r}=\delta_{\mu\nu}.
\label{eq:norm}
\end{align}
In practice this normalization is calculated within a finite-size computational domain, large enough to capture the evanescent field contribution of the QNM~\cite{quasius2}.
An alternative derivation of the normalization scheme is presented
in Ref.~\cite{quasi3}.
The corresponding GF with equal space arguments
within the MNR can then be written as an expansion of the
QNMs~\cite{quasius1,quasius2},
\begin{align}
{\bf G}^{\rm T}({\bf r},{\bf r};\omega)= \sum_{\mu}\frac{\omega^2}{2\tilde\omega_{\mu}(\tilde\omega_{\mu}-\omega)}\tilde{\bf f}_{\mu}({\bf r})\tilde{\bf f}_{\mu}({\bf r}).
\label{G1}
\end{align}
For the gold nanorod of interest, we actually only require
the dominant LSP mode,  and thus obtain a single mode version of
Eq.~(\ref{G1}) as ${\bf G}_c^{\rm T}({\bf r},{\bf r};\omega)= \frac{\omega^2}
{2\tilde\omega_{c}(\tilde\omega_{c}-\omega)}\tilde{\bf f}_{c}({\bf r})\tilde{\bf f}_{c}({\bf r})$.

One problem with the single mode expansion is that
multi-modal contributions near the surface
are not captured. Such effects becomes important when
one is a few nm from a lossy structure interface, causing
a divergence in the LDOS due to quasi-static coupling (Ohmic heating).
However,  boundary effects can
be well described by an image dipole~\cite{nanoptics,localdipole}. 
  Thus, we separate the total scattered
contribution into the contributions of the QNM and image dipole: ${\bf G}_{\rm loc}^{\rm sc} = {\bf G}_{\rm loc}^{\rm qsc} + {\bf
G}_{\rm loc}^{\rm dsc}  $. For cavity positions sufficiently far from
an interface, e.g. 10~nm, then one can safely use
${\bf G}_{\rm loc}^{\rm sc} = {\bf G}_{\rm loc}^{\rm qsc}$ for the LF problem.
For convenience, we define the total enhancement of SE (LDOS) rate as $F_d = F^S_d + F^d_d+ F^0_d$, where
$F_d^{S/d/0}$ are the QNM, image dipole and homogeneous contributions, respectively,  and are
given by $F^{\alpha}_d = \frac{\text{Im}\left\{{\bf n}_d\cdot{\bf G}^{
\alpha}({\bf r}_{d},{\bf r}_d;\omega)\cdot{\bf n}_d\right\}}{\text{Im}
\left\{{\bf n}_d\cdot{\bf G}_\text{0} ({\bf r}_d,{\bf
r}_d;\omega)\cdot {\bf n}_d\right\}}$, with ${\bf G}^{\rm S/d} = {\bf
G}_{\rm loc}^{\rm qsc/dsc}$, and ${\bf G}^{\rm 0} = {\bf G}^{\rm h}_{\rm loc}$,
respectively (${\bf n}_d$ is
the unit vector of polarization).
Here
${\bf G}_{\rm 0}$ is the free space GF,
and $\text{Im} [{\bf n}_d\cdot{\bf G}_0({\bf r}_d,{\bf
r}_d;\omega)\cdot{\bf n}_d]= \omega^3/(6\pi c^3)$.

Given sufficient computational resources, it has been shown that FDTD works very well when dealing with the LFP using a grid size of $(\frac{4\pi r_c^3}{3})^{1/3}$~\cite{finitefdtd}; the SE rate enhancement can be
obtained by injecting an electric dipole at the center of the RC,
and computing the total response of the dipole. This yields the total LF SE emission rate from FDTD, \begin{align}
F_{d}^{\rm FDTD}({\bf r}_a,\omega)
= \frac{\text{Im}\left\{{\bf n}_d\cdot{\bf G}^{\rm FDTD}({\bf r}_{d},{\bf r}_d;\omega)\cdot{\bf n}_d\right\}}{\text{Im}
\left\{{\bf n}_d\cdot{\bf G}_0
({\bf r}_d,{\bf r}_d;\omega)\cdot
{\bf n}_d\right\}}.
\label{ESE1}
\end{align}
Our  goal is to derive $F_d$ analytically, and  to first confirm its accuracy against a brute-force numerical simulation of $F_{d}^{\rm FDTD}$. Then we use our analytical approach to explore some new physics related to LF effects with metallic nanorods.

\begin{figure}
\centering\includegraphics[trim=1.18cm 2.3cm 1.9cm 4.1cm, clip=true,width=.90\columnwidth]{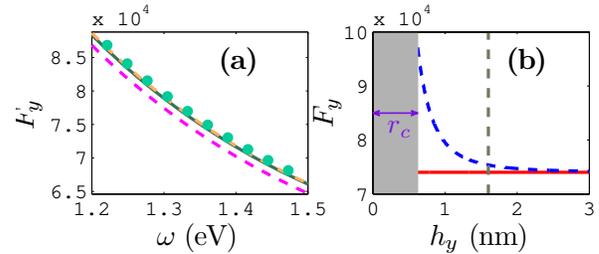}
\caption{\footnotesize{LF SE effect for $r_c=0.62~$nm. (a) Enhancement at 1.5~nm away from the top surface as is shown in Fig.~\ref{f1}(c): ~magenta dashed/green solid/orange dashed/cyan circles/ are $F_y^0$/$F_y^{\rm FDTD}$/$F_y$ iva Eq.~(\ref{dipole})/$F_y$ but using $F_y^d$ given in ~\protect\cite{localdipole}, respectively; (b) enhancement at $\omega_c$ versus the distance $h_y$ (from the top surface to the center of the RC): the red solid/blue dashed are $F_y^0$ and the total calculated via Eq.~(\ref{local}) respectively; the vertical dashed line is the regime of (a).}}
\label{f3}
\vspace{-0.1cm}
\end{figure}

With a RC, whose radius is much less than the scale of the nanorod (for which its effect on the QNM is negligible), then the QNM contribution is given by~\cite{local2}
\begin{align}
{\bf G}_{\rm loc}^{\rm qsc}({\bf r}_d,{\bf r}_d;\omega) = \left(\frac{3\varepsilon}{2\varepsilon + \varepsilon_c}\right)^2{\bf G}_c^{\rm T}({\bf r}_d,{\bf r}_d;\omega),
\label{qsc}
\end{align}
which is obtained analytically from Eq.~(\ref{G1}) with the normalized QNM
$\tilde{\bf f}_c$. So numerically, one must first compute the QNM, but then, unlike FDTD,
 ${\bf G}_{\rm loc}^{\rm qsc}$ is known for all spatial points within the scattering geometry (MNR).
Furthermore,
using a quasistatic approximation-the variation of electric field produced by the image dipole is negligible over the region of the RC, the effect of image dipole  is given by~\cite{nanoptics,comm1}
\begin{align}
{\bf G}_{\rm loc}^{\rm dsc}({\bf r}_d,{\bf r}_d;\omega) = \frac{3\varepsilon}{2\varepsilon + \varepsilon_c}{\bf G}_d({\bf r}_d,-{\bf r}_d;\omega),
\label{Gdsc}
\end{align}
with the image dipole GF  
${\bf G}_d({\bf r},-{\bf
r}_d;\omega)=\pm\frac{\varepsilon_b-\varepsilon}{4\pi\varepsilon(\varepsilon_b+\varepsilon)|{\bf r}_d+{\bf
r}|^3}(\frac{3({\bf r}+{\bf r}_d)({\bf r}+{\bf r}_d)}{|{\bf r}+{\bf
r}_d|^2}-{\bf
1})$, for any position ${\bf r}$
inside the lossy layer ($\varepsilon$), where $\pm$  correspond to the dipole polarized
normal and parallel to the interface, respectively; 
 and
{\bf 1} is the unit dyadic.
Below, we will concentrate on a $y$-polarized electric dipole near the top surface, so the $+$ sign will be used in Eq.~(\ref{Gdsc}), and the enhanced
SE rate from the image dipole is given by
\begin{align}
F_y^d=
9\varepsilon_B\left(\frac{c}{2y_d\omega}\right)^3{\rm
Im}\left [\frac{(\varepsilon_b-\varepsilon)}{(\varepsilon_b+\varepsilon)
(\varepsilon_c+2\varepsilon)}\right ],
\label{dipole}
\end{align}
where $y_d$ is the distance from the center of the cavity 
 to the flat surface. 
 A similar image dipole method has been used recently~\cite{localdipole} to account for the boundary scattering effect of the two layer dielectric structure, though we find that our method is more accurate~as is shown in Fig.~\ref{f3}(a).

We now have, in analytic form,  the total LF GF for the RC inside the lossy MNR,  
\begin{align}
{\bf G}_{\rm loc}^{\rm anal} = {\bf G}_{\rm loc}^{\rm h} + {\bf G}_{\rm loc}^{\rm qsc}
+{\bf G}_{\rm loc}^{\rm dsc},
\label{local}
\end{align}
which separates the contributions into ($i$) a homogeneous contribution, $(ii)$ a QNM contribution, and $(iii)$ a surface contribution. Subsequently,
we have an analytical prescription for obtaining
the total local-field $F_y$.

\begin{figure}[h]
\centering\includegraphics[width=.98\columnwidth]{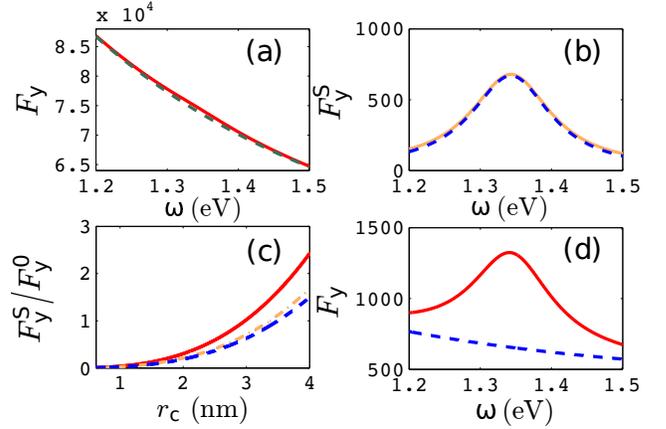} 
\caption{\footnotesize{LF effect: a $y$-polarized dipole at 0.5~nm away from the center of the nanorod as is shown in Fig.~\ref{f1}(b); $r_c = 0.62~$nm for (a-b), and $r_c = 3~$nm for
(d). (a) Full FDTD dipole calculations of $F_y^{\rm FDTD}$ (red solid), and the analytical result of $F_y^0$ (green dashed); (b) Difference between $F_y^{\rm FDTD}$ and $F_y^0$ (orange solid), and the QNM contribution. (c) Ratio of the $F_y^S$ to the $F_y^0$ versus cavity radius: on resonance ($\omega=\omega_c$, red solid),
redshifted ($\omega = 1.3~$eV, orange dot-dashed) 
, and blueshifted ($\omega = 1.4~$eV, blue dashed). (d) $F_y$ (red solid) and the
$F_y^0$ (blue dashed).}}
\label{f2}
\end{figure}

In Fig.~\ref{f1}(b/c)  we show the spatial profile
of the dipole mode of the LSP for the MNR with resonance around 1.34~eV~\cite{quasius2}, and we are
interested in the frequency regime around the LSP. The FDTD~\cite{fdtd} calculation of the homogeneous contribution is described in Ref.~\cite{finitefdtd},
and it can also be calculated
analytically for a spherical RC~\cite{colestrongcoupling,spherical}. 
In Fig.~\ref{f2}(a), the homogeneous contribution to the enhanced SE rate (i.e., $F_y^0$) is  shown  by the green dashed line for a cavity with radius
$r_c=0.62~$nm and refractive index $n_c = 3$ (we use this value for the rest of our paper, but it can easily be changed). 
The FDTD calculation of the total enhancement $F_y^{\rm FDTD}$ (red solid) for the same cavity located 0.5~nm away from the center of
the nanorod along the $y$-axis 
(Fig.~\ref{f1}(b)) shows a  bump around the LSP; 
it is found that the depolarization
effect of the image dipole, $F^d_y$, is negligible at this position.
Due to the position of the resonance bump in $F_y$, we expect it should be an
indication of the LSP. An independent calculation of the scattering
contribution via the QNM using Eq.~(\ref{qsc}) (blue dashed line in
Fig.~\ref{f2}(b)) indeed shows excellent agreement with the difference
between the total and homogeneous contribution of the enhancement,
$F_y^{\rm FDTD}-F_y^0$.

In Fig.~\ref{f2}(c) we show the ratio of $F^S_y$ to $F^0_y$ as a function of the radius of the
RC. It can be seen as $r_c$ increases, the
QNM contribution of the LSP begins to dominate over the homogeneous
contribution.  Figure~\ref{f2}(d) shows  both $F_y$ and $F_y^0$
for a RC with radius $r_c = 3~$nm; the contribution of the LSP
is now larger than  the homogeneous one, and  it also shows a strong
asymmetry around the LSP resonance.

The above observation shows that our formulation, Eq.~(\ref{local}),
 works
well for computing the LF effect, when the RC is at the center of the rod. But for a truly robust technique, it should also work near a surface. Thus in Fig.~\ref{f3}(a), we show
the enhanced SE rate for a RC inside the nanorod but only 1.5~nm away
from the top surface along the $y$-axis (Fig.~\ref{f1}(c)): the green solid line is the FDTD
result $F^{\rm FDTD}_y$, and the magenta dashed is $F^0_y$, which is seen to underestimate the total SE enhancement. Detailed calculation shows that the QNM contribution is now less
than $0.1\%$, and this can be explained by the mode profile of the
QNM, for which the nodal lines lay around both
ends of the nanorod. The orange dashed line in Fig.~\ref{f3}(a) is calculated
with Eq.~(\ref{local}), and once again Eq.~(\ref{local}) agrees well with full FDTD calculation, where now it is important to include the scattering contribution from the image dipole. As shown in Fig.~\ref{f3}(b), the contribution of the image dipole begins to dominate over the homogeneous contribution when the separation between the cavity and the top surface comes into the  sub-nm regime, although quantum tunneling effects near the surface may well also become important here.

Finally, since it is known that MNRs exhibit a large amount of quenching, we have also computed the output $\beta$-factor, which is the probability that a single photon will be emitter into an output radiation mode, i.e., away from the MNR. For a 2 nm RC at the MNR center position, the $\beta$-factor is around 9\%, which is certainly large enough to get light in and out with reasonable efficiencies.

In summary, we have studied the LF problem for a finite-size lossy structure. Using a QNM technique, we have developed
a semi-analytical formulation to compute the LF effect for a finite-size emitter inside a MNR.
The accuracy of this approach is first confirmed by comparing with
an established FDTD
technique. The QNM LF approach enabled us to identify when one gets a significant contribution to the  LF from the LSP resonance.
We have also introduced an image dipole technique to obtain surface contributions for RCs that are near lossy surface. These LF models are of fundamental importance in quantum optics and could have applications for  LSP lasing and spasing.

\section*{Funding Information}
Natural Sciences and Engineering Research Council of Canada.



\begin{thebibliography}{99}
\bibitem{comm1}
The electric field at the positioin of the dipole by the image dipole is $E_y({\bf r}_d) = \frac{p}{2\varepsilon_0\pi}\frac{\varepsilon_b-\varepsilon}{\varepsilon_b+\varepsilon}\frac{3}{(2h)^3(\varepsilon_c+2\varepsilon)}$ with $p$ the dipole moment and $\varepsilon_0$  the permittivity of free space.
\end{thebibliography}
\end{document}